\documentclass[preprint,floats,floatfix,amsmath,amssymb,nofootinbib,longbibliography]{revtex4-1}
\usepackage[hyperindex]{hyperref}
\usepackage{amsfonts}
\usepackage{graphicx}
\usepackage{dcolumn}
\usepackage{bm}
\usepackage{natbib}
\renewcommand{\a}{\approx}
\newcommand{\p}{\partial}
\newcommand{\s}[1]{\scriptscriptstyle (#1)}

\renewcommand{\sc}{\scriptscriptstyle}

\begin{document}
	
\title{The description of gravitational waves in geometric scalar gravity}

\author{J\'unior Diniz Toniato}
\email{junior.toniato@ufes.br}

\affiliation{Center for Astrophysics and Cosmology \& PPGCosmo, Federal University of Esp\'irito Santo - Brazil}

\begin{abstract}
It is investigated the gravitational waves phenomena in the geometric scalar theory of gravity (GSG), a class of theories such that gravity is described by a single scalar field. The associated physical metric describing the spacetime is constructed from a disformal transformation of Minkowski geometry. In this theory, a weak field approximation gives rise to a description similar to that one obtained in general relativity, with the gravitational waves propagating at the same speed as the light, although they have a characteristic longitudinal polarization mode, besides others modes that are observer dependent. We also analyze the energy carried by the gravitational waves as well as how their emission affects the orbital period of a binary system. Observational data coming from Hulse and Taylor binary pulsar is then used to constraint the theory parameter.
\end{abstract}

\date{\today}

\maketitle

\begin{sloppypar}

\section{Introduction}\label{intro}
Although general relativity (GR) has been a very successful gravitational theory during the last century, many proposals for modification of Einstein original formulation appeared in the literature over the past decades. Most of these ideas come up within the cosmological scenario, where GR only works if unknown components, like dark matter or dark energy, are introduced. Such alternative descriptions are basically variations of Einstein's theory, either assuming most general Lagrangians for the gravitational field or adding new fields together with the metric.

Unlike those variations of GR, it was recently proposed a theory of gravity in the realm of purely scalar theories, introducing some crucial modifications from the previous attempts that took place before the emergence of GR \cite{Novello:2012wr}. It represents the gravitational field with a single scalar function $\Phi$, that obeys a non-linear dynamics.\footnote{The non linearity of the field must be specifically in the kinetic term of the Lagrangian, namely $w$\,. Thus, the Lagrangian density of the scalar field can be described as $L=F_1(\Phi,w)w+F_2(\Phi)$\,, with the condition that $F_1$ can not be a constant.} Interaction with matter fields is given only trough a minimal coupling to the physical metric $q_{\mu\nu}$, constructed from a disformal transformation of a auxiliary and flat metric $\gamma_{\mu\nu}$, namely
\begin{equation}\label{qmn}
q_{\mu\nu}=A(\Phi,w)\,\gamma_{\mu\nu}+\frac{B(\Phi,w)}{w}\,\p_\mu\Phi\,\p_\nu\Phi\,,
\end{equation}
with,
\begin{equation}
w=\gamma^{\mu\nu}\p_{\mu}\Phi\,\p_{\nu}\Phi\,,\label{w1}
\end{equation}
and the short notation $\p_\mu=\p/\p x^\mu$. Disformal transformations in gravity has been discussed early by Bekenstein \cite{Bekenstein1993a} and the role of disformal couplings in gravitational theories have attracted interest recently, as in Refs. \cite{Brax2018a,Gannouji2018,Hohmann2019}. Moreover, general ways to deform the spacetime has been considered to enlighten many gravitational problems and metric disformations are also embedded in this class of transformations (see \cite{Capozziello2008b} and the references therein). In this sense, the structure in \eqref{qmn} can be seen as a natural way to introduce a general metric structure through a scalar field.


A complete theory can only be set if one defines the functions $A$ and $B$, and also the Lagrangian of the scalar field. Then, a field equation, characterizing the theory, can be derived. We refer to this class of gravitational theories as geometric scalar gravity (GSG). In early communications on GSG, it was explored a specific set of those functions defining the theory, which shows that it is possible to go further in representing the gravitational field as a single scalar, giving realistic descriptions of the solar system and cosmology \cite{Novello:2012wr,Bittencourt:2014oua}. An analysis of GSG within the so called parametrized post-Newtonian formalism was also made and, although the theory is not covered by the formalism, a limited situation indicate a good agreement with the observations \cite{Bittencourt:2016smd}. Intending to improve the understanding of how GSG deals with gravitational interaction, the present work develops the theoretical description and characterization of gravitational waves (GW).

The direct detections of GW by LIGO and Virgo collaborations initiated a new era of testing gravitational theories. It enables to construct constraints over a series of theoretical mechanisms associated with GW's physics, but crucial point relies on the observed waveform and how a theory can reproduce it \cite{Yunes2016}. Notwithstanding, this is not the scope of this work. We are mainly focused in analyzing the GW fundamentals on the perspective of GSG, studying their propagation, polarization modes and defining an appropriated tensor to describe the energy and momentum carried by the waves. The velocity of propagation of GW has been measured with good precision indeed, but this data does not constraint GSG once the gravitational signal travels in vacuum with the same speed of light, as it will be shown later. However, the theory can be constrained by observational data from pulsars through its prediction for the orbital variation of a binary system that should be caused by the loss of energy due to gravitational radiation.

The paper is organized as follows. In section \ref{1} is presented a brief overview of GSG in order to introduce to the reader the main features of this theory. The following section describes the theory's weak field approximation. In section \ref{3} the study of the propagation and vibration modes associated to gravitational waves is made. The definition of a energy-momentum tensor for the linear waves is treated in section \ref{egw}. Generation of waves, including the computation of the orbital variation of binary systems due to the emission of GW, is discussed in section \ref{5} and the last section presents our concluding remarks. Also, two appendices were introduced in order to complement the middle steps of calculations present in section \ref{5}.

\section{Overview of geometric scalar gravity}\label{1}
GSG is a class of gravitational theories which identifies the gravitational field to a single real scalar function $\Phi$, satisfying a non-linear dynamic described by the action
\begin{equation}
S_\Phi=\int{\sqrt{-\gamma}\,L(\Phi,w)d^4x}\,,
\end{equation}
where $\gamma$ is the determinant of the Minkowski metric and $w$ is defined in eq.\,\eqref{w1}. Metric signature convention is $(+,-,-,-)$. The physical metric is constructed from the gravitational field according to the expression \eqref{qmn} and its contravariant form is obtained from the definition of an inverse metric, $q^{\mu\alpha}q_{\alpha\nu}=\delta^\mu_\nu$, namely,
\begin{equation}
q^{\mu\nu}=\alpha(\Phi,w)\,\gamma^{\mu\nu}+\frac{\beta(\Phi,w)}{w}\,\gamma^{\mu\alpha}\gamma^{\nu\beta}\p_{\alpha}\Phi\p_{\beta}\Phi\,,\label{Qm}
\end{equation}
where,
\begin{equation}
A=\frac{1}{\alpha}\quad \mbox{and}\quad B=-\frac{\beta}{\alpha(\alpha+\beta)}\,.
\end{equation}

In order to describe the interaction of the scalar gravitational field with matter, GSG makes the fundamental hypothesis, according to Einstein's proposal, that gravity is a geometric phenomenon. Thus, it is assumed that the interaction with $\Phi$ is given only through a minimal coupling with the gravitational metric $q_{\mu\nu}$.
~The matter action in GSG is then described as
\begin{equation}
S_{m} = \int{\sqrt{- q}\, L_{m} \,d^4x}\,.
\end{equation}

A complete theory should specify the metric's functions $A$ and $B$ together with the Lagrangian of the scalar field $L$, in order to be possible to derive its field equation. Up to now in the literature, it has been explored the case in which the following choice is made,
\begin{align}
	& \alpha= e^{-2\Phi}\,,\label{set1}\\[1ex]
	& \beta=\frac{(\alpha-1)(\alpha-9)}{4}\,,\label{set2}\\[1ex]
	& L=V(\Phi)\,w\,,\label{set3}
\end{align}
with
\begin{equation}\label{v}
V=\frac{(\alpha-3)^{2}}{4\alpha^{3}}.
\end{equation}
Using the standard definition of the energy momentum tensor in terms of a metric structure, we set
\begin{equation}
T_{\mu\nu}\equiv \frac{2}{\sqrt{- q}}\,\frac{\delta( \sqrt{- q} \, L_{m})}{\delta q^{\mu\nu}}\,.
\end{equation}
Then, the dynamics of the scalar field is described by the equation
\begin{equation}\label{eqgsg}
\sqrt{{V}} \, \square\Phi= \kappa \, \chi\,,
\end{equation}
where the $\Box$ indicates the d'Alembertian operator constructed with the physical metric $q_{\mu\nu}$, $\kappa$ is a coupling constant and the source term $\chi$ is provided by
\begin{equation}\label{chi}
\chi=\,-\,\frac{1}{2}\left[\,T +\left(2 -\frac{V'}{2V}\right)E +C^\lambda_{~\,;\lambda}\right]\,,
\end{equation}
where `$\,;\,$' indicates a covariant derivative with respect to the physical metric, $V'=dV/d\Phi$ and
\begin{align}
	T &=T^{\mu\nu}q_{\mu\nu} \label{t}\,,\\[1ex]
	E &= \frac{T^{\mu\nu} \, \partial_{\mu}\Phi \, \partial_{\nu}\Phi}{\Omega}\,,\label{e}\\[1ex]
	C^{\lambda} &=\frac{\beta}{\alpha\Omega} \, \left( T^{\lambda\mu} - E \, q^{\lambda\mu} \right) \, \partial_{\mu}\Phi\,,\label{c}\\[1ex]
	\Omega&=q^{\mu\nu}\p_\mu\Phi \p_\nu\Phi.
\end{align}

The choices made in \eqref{set1}-\eqref{set3} are such that the resulting theory satisfies the Newtonian limit, the classical gravitational tests and the spherically symmetric vacuum solution is given by the Schwarzschild geometry. Moreover, in the absence of any matter fields, $\Phi$ is a free wave propagating in the metric $q_{\mu\nu}$ \cite{Goulart:2011cb}. More details concerning the fundamentals of GSG and how this specific model can successfully describe the solar system physics and cosmology can be found in \cite{Novello:2012wr,Bittencourt:2014oua,Bittencourt:2016smd}. In the present work we will consider only this model. To work with different functions $\alpha$, $\beta$ and $L$, all the process of constructing the field equation of the theory has to be redone, as well as it should be checked the feasibility of the new theory.

\section{Weak field approximation}\label{2}
To discuss linear gravitational waves we should consider an isolated system, distant from any source, embedded in a homogenous and isotropic universe. At a particular moment of time and specific distance from the isolated system, the background metric can be transformed to assume a flat Minkowskian form, resulting in a geometric structure given by,
\begin{equation}\label{wfa}
q_{\mu\nu}=\eta_{\mu\nu}+h_{\mu\nu}\,,
\end{equation}
where $\eta_{\mu\nu}=\mbox{diag}(1,-1,-1,-1)$ and $h_{\mu\nu}$ represents the first order perturbations.

In this sense, the weak-field approximation of GSG consists in a small deviation of a cosmological solution $\phi_{0}$. Thus, we set
\begin{equation}\label{pertphi}
\Phi\a \phi_{0} + \phi\,, \quad \mbox{with} \quad  |\phi|\ll 1.
\end{equation}
In order to construct the geometric structure as in \eqref{wfa}, for simplicity, we start with a coordinate system $\tilde{x}^{\mu}$, where the auxiliary metric $\gamma_{\mu\nu}$ assumes the usual diagonal form indicated as $\eta_{\mu\nu}$, and we expand the kinetic term and the metric coefficients as follows,
\begin{align}\label{coef}
	& w\approx w_{0}+w_{1}\,,\\[1ex]
	& w_{0}=(\p_{\tilde{0}}\phi_{0})^{2}\,,\quad w_{1}=2(\p_{\tilde{0}}\phi_{0})(\p_{\tilde{0}}\phi)\,\\[1ex]
	& \alpha \a \alpha_{0}(1 - 2\phi), \ \beta \a \beta_{0}- \alpha_{0}(\alpha_{0}-5)\,\phi\,.
\end{align}
The subindex ``$_{0}$'' identifies quantities constructed with $\phi_{0}$ according to basic expressions given in the previous section. The gravitational metric takes the form
\begin{equation}
\tilde{q}^{\,\mu\nu}=\tilde{q}_{0}^{\,\mu\nu}- \tilde{h}^{\mu\nu}\,,
\end{equation}
where,
\begin{equation}
\tilde{q}_{0}^{\,00}=\frac{(\alpha_{0}-3)^{2}}{4}\,, \ \tilde{q}_{0}^{\,0i}=0\,, \ \tilde{q}_{0}^{\,ij}= -\alpha_{0}\delta^{ij}\,.
\end{equation}
and,
\begin{equation}
\tilde{h}^{00}=\alpha_{0}(\alpha_{0}-3)\phi,\ \tilde{h}^{0i}=\frac{\beta_{0}}{\sqrt{w_{0}}}\,\p_{\,\tilde{j}}\phi\,\delta^{ji},\ \tilde{h}^{ij}=-\,2\alpha_{0}\phi\delta^{ij}\,.
\end{equation}
With the following coordinate transformation
\begin{equation}\label{transf}
x^{0}=\frac{2\tilde{x}^{0}}{3-\alpha_{0}}\,, \quad x^{i}=\frac{\tilde{x}^{i}}{\sqrt{\alpha_{0}}}\,,
\end{equation}
the desired structure is achieved,
\begin{equation}\label{pertmetric}
q^{\mu\nu}=\eta^{\mu\nu}-h^{\mu\nu}\,,
\end{equation}
where
\begin{equation}
h^{00}=\frac{4\tilde{h}^{00}}{(\alpha_{0}-3)^{2}}\,, \quad h^{0i}=\frac{2\tilde{h}^{0i}}{\sqrt{\alpha_{0}}(3-\alpha_{0})}\,, \quad h^{ij}=\frac{\tilde{h}^{ij}}{\alpha_{0}}\,.
\end{equation}
In this new coordinate system, $x^{0}$ is equivalente to the cosmological time and $\p_{0}\phi_{0}=H_{0}$ (in units where $c=1$), where $H_{0}$ is the Hubble parameter (please see \cite{Bittencourt:2014oua} for more details on GSG cosmology).\footnote{The index $_0$ in the Hubble parameter is used to indicate a background quantity.} Then, the perturbed metric becomes,
\begin{equation}\label{hs}
h^{00}=\frac{4\alpha_{0}}{\alpha_{0}-3}\,\phi\,, \quad h^{0i}=\frac{\beta_{0}}{\alpha_{0}H_{0}}\p_k\phi\delta^{ki}\,, \quad h^{ij}=-\,2\phi\,\delta^{ij}\,.
\end{equation}

The corresponding covariant expression for \eqref{pertmetric} is obtained from the definition $q_{\mu\alpha}q^{\alpha\nu}=\delta^\mu_\nu$\,. It reads
\begin{equation}\label{pertm}
q_{\mu\nu}= \eta_{\mu\nu} + \,h_{\mu\nu}\,,
\end{equation}
where,
\begin{align}
	&h_{\mu\nu}= \eta_{\mu\alpha}\eta_{\nu\beta}\,h^{\alpha\beta}\,,
\end{align}
Equations \eqref{pertmetric} and \eqref{pertm} shows that in the weak field limit the indices are lowered and raised by the Minkowski background metric.

Note that, the perturbed metric \eqref{hs} can also be derived from the expansion of the exact form given in \eqref{Qm} starting already with the coordinates $x^\mu$, where the auxiliary metric $\gamma^{\mu\nu}$ takes the form
\begin{equation}
\gamma^{\mu\nu}=\left(\frac{1}{\alpha_{0}+\beta_{0}}\,, -\frac{\delta_{ij}}{\alpha_{0}} \right).
\end{equation}
The resulting covariant expression can be written as
\begin{align}
	h^{\mu\nu} =-2\alpha_{0}\phi\,\gamma^{\mu\nu} + \gamma^{\mu\alpha}&\gamma^{\nu\beta} \bigg[\frac{\beta_{0}}{w_{0}}H_{(\alpha}\p_{\beta)}\phi \ + \notag\\
	& + \ \left(\frac{\beta_{1}}{w_{0}}- \frac{b_{0}w_{1}}{w_{0}^{2}} \right)H_{\alpha}H_{\beta}\bigg],
\end{align}
with
\begin{equation}
H_{\mu}=\p_\mu\phi_{0}=(H_{0},0,0,0)\quad\mbox{and}\quad \beta_{1}=\alpha_{0}(\alpha_{0}-5)\phi.
\end{equation}

In reference \cite{Bittencourt:2016smd} a distinct weak field approximation was made where the scalar field was expanded around a vanishing background value. Although consistent, that scheme is not suitable for the description of GW, due to a term $\p_\mu\phi\p_{\nu}\phi/w$ that is present in $h_{\mu\nu}$. Oscillatory solutions would then lead to a singular behavior of the metric, evidencing that the background cosmological scenario can not be neglected.

\subsection{The cosmological backgroung}
Before proceeding in the analysis of GW in GSG, let us clarify important points of the cosmological background described by $\phi_{0}$. To a more detailed discussion about the cosmology in GSG we refer to the reader the analysis present in \cite{Bittencourt:2014oua}. By considering the scalar field as a function of coordinate $\tilde{x}^{0}$ only, the metric arising is of Friedman-Robertson-Walker type with a flat spatial section. The cosmological time is achieved by the time transformation given by the first expression in \eqref{transf} and the scale factor, said $a_0(x^{0})$, is related with the $\phi_{0}$ as follows,
\begin{equation}
a^{2}_{0}=\frac{1}{\alpha_{0}}.
\end{equation}

The dynamical equation \eqref{eqgsg} contains two regimes classified by the term $\sqrt{V_{0}}\propto |\alpha_{0}-3|$, a consequence of the particular choice of the scalar field Lagrangian. The case where $\alpha_{0}<3$ $(a_{0}>1/\sqrt{3})$, with a barotropic fluid as source, describes a eternal universe without singularities. The universe has a bouncing, followed by a early accelerated phase and a final decelerated expansion. The problematic value $\alpha_{0}=3$ is unattainable, in other words, the minimal value of the scale factor $a_{0}$ is always greater than $1/\sqrt{3}$. A distinct behavior occurs for the solutions with barotropic fluids in the region where $\alpha_{0}>3$ $(a_{0}<1/\sqrt{3})$; the universe starts from a initial singularity, it expands to a certain maximum value of the scale factor, smaller than $1/\sqrt{3}$, and then returns to a final singular point. This two regions are then disjoint classes of cosmological solutions.
In the present work, we will consider only the case where
\begin{equation}
\alpha_{0}<3,
\end{equation}
since it represents a class of more realistic descriptions of the universe.

\section{Propagation and polarization of gravitational waves}\label{3}
At the level of the dynamical equation we can consider $\phi_{0}$ as a constant, since its timescale variation is longer compared to the dynamical timescale for the local system. Expanding the left hand side of Eq. \eqref{eqgsg} and neglecting second order terms, one has,
\begin{equation}\label{lhs}
\sqrt{V}\,\Box\Phi \a \sqrt{V_{0}}\,\Box_{\eta}\phi\,,
\end{equation}
where we refer to Minkowskian d'Alembertian operator as $\Box_{\eta}\,.$ Thus, without the presence of sources, one has
\begin{equation}
\Box_{\eta}\phi=0\,. \label{feq1v}
\end{equation}

The perturbed scalar field has oscillatory solutions which propagates at the speed of light. Once the metric is constructed with the field and its first derivatives, such solutions yields oscillations as GW in the geometric structure of the spacetime. Moreover, it is verified that
\begin{equation}
\Box_{\eta}h_{\mu\nu}=0\,,
\end{equation}
thus gravitational waves in GSG does propagate with the speed of light, showing no deviation with respect to GR in this aspect. Consequently, GSG is also supported by the combined data of the GW event GW170817 and the gamma-ray burst GRB 170817A, which constraint the velocity of propagation of GW to be the same as the speed of light within deviations of order $10^{-15}$  \cite{Abbott_2017}.

\subsection{Polarization states}\label{4}
The most general (weak) gravitational wave that any metric theory of gravity is able to predict can contain six modes of polarization. Considering plane null waves propagating in a given direction, these modes are related to tetrad components of the irreducible parts of the Riemann tensor, or the Newmann-Penrose quantities (NPQ): $\Psi_2, \Psi_3, \Psi_4$ and $\Phi_{22}$ ($\Psi_3$ and $\Psi_4$ are complex quantities and each one represents two modes of polarization) \cite{Newman:1961qr}. The others NPQ are negligible by the weak field approximation, or are described in terms of these four ones. 

The linearized dynamical equations of a gravitational theory can automatically vanish some of these NPQ, specifying then the  predicted number of polarization states. For instance, in GR only $\Psi_{4}$ is not identically zero, which characterizes two transversal polarization modes, called ``$+$'' and ``$\times$'' states. In general, the six polarization modes can not be specified in a observer-independent way because of their behavior under Lorentz transformations. Nevertheless, if we restrict our attention to a set of specific observers, which agree with the GW on the frequency and on the direction of propagation, then is possible to make some observer-invariant statements about the NPQ. Such assertions are on the basis of the so called \emph{E(2)-classification} of gravitational theories, introduced in ref.~\cite{Eardley:1974nw}:
\begin{itemize}
	\item \emph{Class II$_6$}: If $\Psi_2 \neq 0$\,, all the standard observers agree with the same nonzero $\Psi_2$ mode, but the presence or absence of the other modes is observer-dependent.
	
	\item \emph{Class III$_5$}: If $\Psi_2=0$\, and $\Psi_3\neq 0$\,, all the standard observers measure the absence of $\Psi_2$ and the presence of $\Psi_3$, but the presence or absence of all other modes is observer dependent.
	
	\item \emph{Class N$_3$}: If $\Psi_2=\Psi_3=0\,, \Psi_4\neq 0$ and $\Phi_{22}\neq 0$\,, this configuration is independent of observer.
	
	\item \emph{Class N$_2$}: If $\Psi_2=\Psi_3=\Psi_2=0$\, and $\Psi_4\neq 0$\,, this configuration is independent of observer.
	
	\item \emph{Class O$_1$}:  If $\Psi_2=\Psi_3=\Psi_4=0$\, and $\Phi_{22}\neq 0$\,, this configuration is independent of observer.
	
	\item \emph{Class O$_0$}:  If $\Psi_2=\Psi_3=\Psi_4=\Phi_{22}=0$\,, this configuration is independent of observer.
\end{itemize}

\begin{figure}[t]
	\begin{center}
		\includegraphics[scale=0.8]{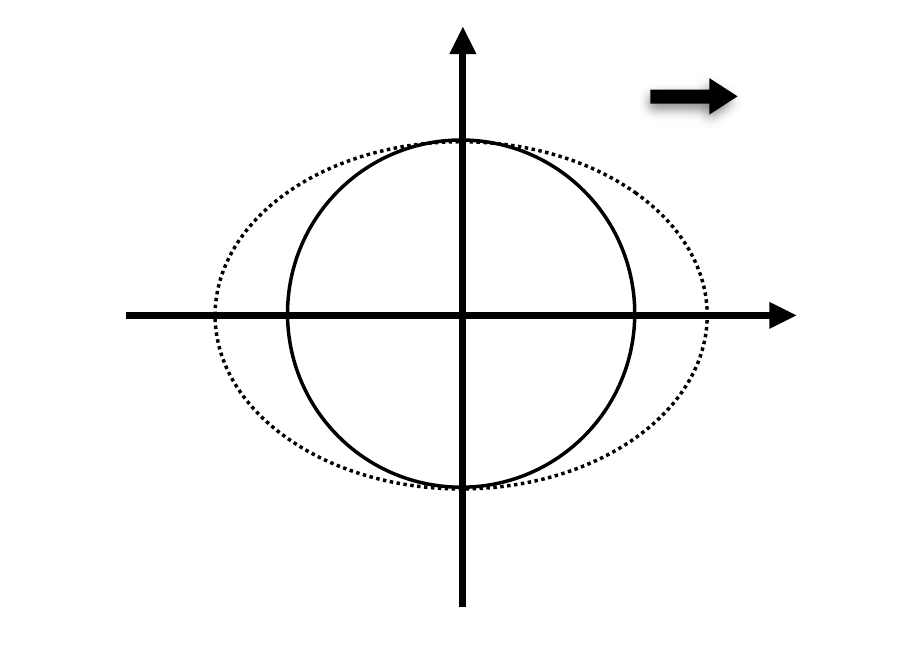}
		\caption{Diagram for the effects of a $\Psi_{2}$-polarized wave when passing through a ring of test particles. The black arrow in the upper right indicates the direction of propagation of the wave.}
		\label{fig}
	\end{center}
\end{figure}

The $E(2)$-classification of GSG is easily obtained by noticing that the Ricci scalar is not identically null. Actually, from the weak field approximation, one has
\begin{equation}
R \a \p_\mu \p_\nu h^{\mu\nu}-\Box_{\eta}h\,,
\end{equation}
with $h=\eta^{\mu\nu}h_{\mu\nu}$\, and, using relations \eqref{hs} together with linearized vacuum field equation \eqref{feq1v}, it is verified that,
\begin{equation}
R\a \frac{2(\alpha_{0}+3)\,\p_t^{2}\phi}{(\alpha_{0}-3)c^2} - \frac{2\beta_{0}\,\p_t^{3}\phi}{\alpha_{0}H_0c^2}\,.
\end{equation}
This result implies $\Psi_2\neq0$ (cf. equation (A4) of \cite{Eardley:1974nw}) and GSG is from the class \emph{II}$_6$\,. This $\Psi_{2}$ represents a pure longitudinal polarization state (see figure \ref{fig}) that is always present in the GW, although other modes can be detected depending on the observer.

GSG belongs to the most general class of theories with respect to the $E(2)$-classification, where is always possible to find an observer that measures all six gravitational wave modes. The authors in \cite{Eardley:1974nw} already pointed out the fact that the number of polarization states predicted by a gravitational theory does not necessarily match the numbers of degrees of freedom inside the theory. They also give an example of this with the so called stratified theories. Other examples of theories also classified as $\emph{II}_{6}$ is the well know $f(R)$ extensions of general relativity \cite{Alves:2010ms,Alves:2009eg,Rizwana:2016qdq}.

Thus, the description of GW by GSG carries a substancial distinction from GR, as it predicts the presence of a longitudinal polarization mode. Up to now, the recent detections of GW can not exclude the existence of any one of the six modes of polarization \cite{LIGOScientific:2019fpa,Abbott:2018lct}. But in the future, with the appropriated network of detectors, with different orientations, this information can be used to restrict gravitational theories.

\section{Energy of the gravitational wave}\label{egw}
In order to associate an energy-momentum tensor to the gravitational waves in GSG we follow a standard procedure, identifying the relation between the second and the first order perturbations of the gravitational field \cite{padmanabhan2010gravitation}. First note that, without approximations, the following relation holds,
\begin{equation}
\Box\Phi= \alpha^{3}V\left(\Box_{\eta}\Phi+\frac{dV}{d\Phi}\frac{w}{2V}\right).
\end{equation}
Thus, taking $\phi \a \phi_{\sc (1)} + \phi_{\sc (2)}$, where the subindexes indicates the order, and computing the second order vacuum field equation, it yields
\begin{equation}\label{fe2}
\Box_{\eta}\phi_{\sc (2)}= -\,\frac{9-\alpha_{0}}{3-\alpha_{0}}\,w_{\sc (2)},
\end{equation}
with $w_{\sc (2)}=\eta^{\mu\nu}\p_{\mu}\phi_{\sc (1)}\p_{\nu}\phi_{\sc (1)}$. The right hand side of this equation contains only the derivatives of the first order field $\phi_{\sc (1)}$, thus it can be interpreted as the source for the second order field generated by the linear waves.

From the general structure of the field equation of GSG, the influence of any energy-momentum tensor enters in the equation of motion uniquely through the quantity $\chi$ [cf. equation \eqref{chi}]. Thus, the energy-momentum tensor of the GW, said $\Theta_{\mu\nu}$, must be consistent with,
\begin{equation}\label{f3}
\chi_{\sc (2)}(\Theta_{\mu\nu})= - \dfrac{9-\alpha_{0}}{2\kappa \alpha_{0}^{3/2}} \,w_{\sc (2)}\,,
\end{equation}
where $\chi_{\sc (2)}(\Theta_{\mu\nu})$ means the second order approximation of the source term calculated with the energy-momentum tensor of the gravitational field $\Theta^{\mu\nu}$, instead of $T^{\mu\nu}$. Therefore, we write
\begin{equation}
\sqrt{V_{0}}\,\Box_{\eta}\phi_{\sc (2)}=\kappa\chi_{\sc (2)}(\Theta_{\mu\nu})\,,
\end{equation}
which has the same general structure of GSG's field equation.

To describe the energy and momentum carried by the linear waves, the second-order approximation of $\Theta_{\mu\nu}$ must be quadratic in the first derivatives of $\phi_{\sc (1)}$. This lead us to a specific form for it,
\begin{equation}\label{tet}
\Theta_{{\s{2}} \mu\nu}= \frac{1}{\kappa}\left(\sigma w_{\sc (2)}\eta_{\mu\nu} +\lambda\,\p_\mu\phi_{\sc (1)}\,\p_\nu\phi_{\sc (1)}\right)\,,
\end{equation}
with $\sigma$ and $\lambda$ being arbitrary constants. The condition \eqref{f3} returns the relation
\begin{equation}\label{c1}
\sigma\left(\frac{9-5\alpha_{0}}{3-\alpha_{0}}\right) - \lambda\left(\frac{2\alpha_{0}}{3-\alpha_{0}}\right)=\frac{9-\alpha_{0}}{\alpha_{0}^{3/2}}\,.
\end{equation}
Any tensor, described like in Eq. \eqref{tet} and satisfying the above relation, can be used as the energy-momentum tensor of the linear GW in GSG. This ambiguity already appeared in reference \cite{Novello:2013tka}, where the authors show how to construct the energy-momentum tensor of the gravitational field in GSG, without using approximate methods. In that occasion, they fixed the functions defining the energy tensor by requiring that $\Theta_{\mu\nu}$ can be derived from the Lagrangian. As expected, their results are consistent with the relation above and are recovered (inside the approximation method) if $\sigma=2\sqrt{V_0}$ and $\lambda=-4\sqrt{V_0}$. In what follows we will proceed with the generic expression for $\Theta_{\mu\nu}$ and look for a specific example, the orbital variations in binary systems, to see how this ambiguity can influence in a observed phenomenon.

\section{Orbital variation of a binary system}\label{5}
This section focus on deriving an expression for the orbital variation of a binary system, due to the emission of gravitational waves, as it is predicted by GSG. In order to obtain the energy rate emitted by the system one should consider the influence of the source into the dynamics in the linear approximation. Since the left hand side of Eq. \eqref{eqgsg} reduces to a Minkowskian d'Alembertian when linearized (c.f. eq.\eqref{lhs}), from the method of Green functions, we immediately write down the general solution as,
\begin{equation}\label{gensol}
\phi(t,\vec{x})= \dfrac{\kappa}{4\pi\sqrt{V_{0}}}\int{\frac{\chi_{L}(t_{r},\vec{x}')}{|\vec{x}-\vec{x}'|}\,d^3x'}\,,
\end{equation}
where $\chi_{L}$ attends to the first order approximation of the source term [cf.\,\eqref{chi}] and $\,t_r=t-r/c$, with $r=|\vec{x}-\vec{x}'|$, is the retarded time.

By considering that the source is far away from the point where we calculate the scalar field $(R \gg r'$, where $R = |\vec{x}|$, and $r'= |\vec{x}'|$ is the typical distances between the source's components$)$, it is possible to make a multipole expansion\footnote{Although symbol $R$ was previously used as the Ricci scalar in section \ref{3}, we draw the reader's attention that in this section it is indicating the point where the gravitational field is being calculated.}. 
Further assuming that the typical velocities of the source components are non relativistic, it is also possible to expand the time dependent terms of the integrand in a Taylor series. For our purpose here it is sufficient to take only the first term of this expansion. Thus, one has
\begin{equation}\label{mod}
\phi(t,\vec{x})\a \frac{\kappa}{4\pi \sqrt{V_{0}}R}\int{\chi_{L}(t_{R},\vec{x}')\,d^3x'}\,,
\end{equation}
where $t_R=t-R/c$ and we have neglected terms of order $1/R^{2}$.


Most of terms in the above integration contains the scalar field $\phi$\,, explicitly. To solve them, we have to expand these terms using the correspondent post-Newtonian approximation of the field in the near-zone region \cite{will1993theory}. However, to keep the final result up to order $G^{2}/c^{4}$, it is only necessary the Newtonian approximation of the near-zone scalar field, namely $\Phi_{N}$. By the viral theorem, we know that, for slow motions, $v^{2}\sim GM/R$, where $v$, $M$ and $R$ are the typical velocity, mass and distances in the source's components, respectively. Thus, $\Phi_{N}\sim v^{2}/c^2$, $\p_{0}\Phi_{N} \sim v^{3}/c^3$ and $\p_{i}\Phi_{N}\sim v^{2}/c^2$ (see appendix \ref{apt} for more details). The energy-momentum tensor also depends on source velocities with $T^{0i}\sim v/c$ and $T^{ij}\sim v^{2}/c^{2}$\,. Thus, keeping terms up to order $v^2/c^2$ (since $\kappa\sim v^{2}/c^{2}$) and using the perturbed metric expressions in \eqref{hs}, one gets
\begin{align}
	T^{\mu\nu}\eta_{\mu\nu} &\a T^{00}-T^{ij}\delta_{ij}\,,\\[2ex]
	T^{\mu\nu}h_{\mu\nu} &\a \frac{4\alpha_{0}}{\alpha_{0}-3}\,\Phi_N\,T\,^{00}\,,\label{tl}\\[2ex]
	\left(2-\tfrac{V'}{2V}\right)E &\a  \ - \ \frac{(3+\alpha_0)}{(3-\alpha_{0})}\dfrac{1}{w_N}\Big(T^{\,00}(\p_{0}\Phi_{N})^{2} \ + \notag\\
	& \quad \ + \ 2T^{\,0i}\p_{0}\Phi_{N}\p_{i}\Phi_{N} +T^{ij}\p_i\Phi_N \p_j\Phi_N \Big),\label{el}\\[2ex]
	C^0 &\a - \ \frac{\beta_{0}}{\alpha_0 w_N}\bigg[\frac{E\,\p_0\Phi_{N}}{w_{N}} \ + \notag\\
	& \quad + \ \left(1+\tfrac{9-\alpha_0^{2}}{2\alpha_0}\,\Phi_{N}\right)(T^{00}\p_0\Phi_{N}+T^{0i}\p_{i}\Phi_{N})\bigg] \label{cl}
\end{align}
The $C^i_{~,i}$ term does not contribute by Gauss law. Also, to derive the above expressions we take into account that
\begin{align}
	&w_{N}\a \eta^{ij}\p_{i}\Phi_{N}\p_{j}\Phi_{N}\sim v^{4}/c^{4}.
\end{align}

Specifying the source for the case of a binary system, we have
\begin{align}
	T^{00} &=\sum_{n}\,m_nc^2\left(1 + \frac{v^2_n}{2c^2}+\Phi_N\right)\delta^3(\vec{x}-\vec{x}_n) + {\cal O}(v^4/c^4)\,,\label{bitensor}\\[2ex]
	T^{0i} &=\sum_{n}\,m_n c\, v_n^i\,\delta^3(\vec{x}-\vec{x}_n)+ {\cal O}(v^3/c^3)\,,\label{bitensor2}\\[2ex]
	T^{ij} &= \sum_{n}\,m_n v^i_n v^j_n\,\delta^3(\vec{x}-\vec{x}_n)+ {\cal O}(v^4/c^4)\,,\label{bitensor3}
\end{align}
where summation is over the two particles of the system, i.e. $n=1,2$. With these expressions, all the integrals in \eqref{mod} can be analytically calculated (more details in Appendix \ref{b}) to give
\begin{align}
	\phi\a& \ \dfrac{(\alpha_0-3)}{2\alpha_0}\dfrac{G}{c^{4}R}\bigg[X_0(\dot{r}^{2}+r\ddot{r}) + 4G\,\dfrac{m_1m_2}{r} \ +\notag\\[1ex]
	& \qquad \ + \ \dfrac{(\alpha_0+3)}{(\alpha_0-3)}\,M\dot{r}^{2} + \dfrac{Y_0}{G}\,(2r\dot{r}^{4}+3r^{2}\dot{r}^{2}\ddot{r}) \ + \nonumber\\[1ex]
	&+ \, \dfrac{\beta_0}{\alpha_0-3} \dfrac{m_1m_2}{GM^{2}}\,(4r^{3}\dot{r}^{2}\dot{\theta}^{2}+r^{4}\dot{\theta}^{2}\ddot{r}+ 2r^{4}\dot{r}\dot{\theta}\ddot{\theta} )\bigg] +\emph{C},\label{phi}
\end{align}
where the dot indicates a derivative with respect to retarded time, $\emph{C}$ attends to constant terms that does not contribute to the radiation,  $G$ is the Newtonian gravitational constant as measured today (see Appendix \ref{apt}) and the notation was shortened by the definitions below,
\begin{align}
	&X_0\equiv\dfrac{m_1m_2}{M}+\dfrac{\beta_0}{\alpha_0}\dfrac{(m_1^{2}+m_2^{2})}{M}-\left(\dfrac{9-\alpha_0^{2}}{2\alpha_0}\right)M,\\[1ex]
	&Y_0\equiv\dfrac{\beta_0}{\alpha_0-3}\left(\dfrac{m_1m_2}{M^{2}}+\dfrac{8\alpha_0(m_1^{5}+m_2^{5})}{m_1m_2M^{3}}\right).
\end{align}
Also, we are adopting the usual center of mass notation such that,
\begin{equation}\label{rs}
\vec{r}_1=\frac{m_2}{M}\,\vec{r}\quad \mbox{and} \quad \vec{r}_2=- \ \frac{m_1}{M}\,\vec{r}\,,
\end{equation}
with $\vec{r}=\vec{r}_1-\vec{r}_2$ and $M=m_1+m_2$\,. 


Once we are dealing with a binary system as the source of the gravitational field, we can use the Keplerian orbital parameters to simplify the above expression \cite{straumann2013general}. The distance between the two masses are,
\begin{equation}\label{r}
r= \frac{a(1-e^2)}{1+e \cos{\theta}}\,,
\end{equation}
where $a$ is the semimajor axis and $e$ is the eccentricity of the orbit. They are related with the total energy $E$ and the angular momentum $L$ by
\begin{equation}\label{ae}
a= -\frac{G\,m_1 m_2}{2E}\,, \qquad e^2=1+2\,\frac{EL^2}{G^2}\frac{M}{(m_1 m_2)^3}\,,
\end{equation}
with $E<0$. The fact that $L=(m_1 m_2/M)r^2 \dot{\theta}\,$, allow us to derive the following relation,
\begin{equation}\label{teta}
\dot{\theta}=\sqrt{\frac{GM}{a^3(1-e^2)^3}}\,(1+e\cos{\theta})^2\,.
\end{equation}
Then, in \eqref{phi}, all time derivatives can be expressed in terms of $\theta$, yielding
\begin{align}
	\phi\a& \ \dfrac{(\alpha_0-3)G^{2}M}{2a(1-e^{2})\alpha_0c^{4}R}\bigg[\Big(X_0 + \dfrac{m_1m_2}{M}\Big(4+\dfrac{\beta_0}{\alpha_0-3}\Big)\Big)e\cos\theta - \notag\\[1ex]
	& \ - \dfrac{(\alpha_0+3)}{(\alpha_0-3)}\,Me^{2}\cos^{2}\theta + Y_0Me^{3}\sin^{3}\theta\,\Big(\dfrac{2e\sin\theta}{1+e\cos\theta} \ + \notag\\[1ex]
	& \ + \ 3\cos\theta\Big) + \dfrac{\beta_0\,m_1m_2}{(\alpha_0-3)M}\,e^{2}\cos^{2}\theta(2+e\cos\theta)\bigg] +\emph{C},
\end{align}

To calculate the energy-flux that is carried off by GW we use the gravitational energy-momentum tensor presented in the previous section. The flux in the radial direction will be $c\,\Theta^{0r}$ thus, the energy radiated per unit time that is passing through a sphere of radius $R$, is given by
\begin{equation}\label{erad}
\frac{dE}{dt}=\dfrac{2\alpha_0^{5/2}\lambda}{(3-\alpha_0)^{2}}\frac{c^{3}R^{2}}{G}\,\dot{\phi}^{2},
\end{equation}
where we have used the fact that
\begin{equation}
\p_0\phi=\frac{1}{c}\,\p_{t_R}\phi\quad \mbox{and} \quad \p_i\phi=-\frac{x^i}{cR}\,\p_{t_R}\phi +{\cal O}\left(\frac{1}{R^2}\right).
\end{equation}

At this point, we go further in the approximation scheme in order to get a more treatable expression for the rate of energy loss. Let us consider that the background field is too small, i.e. $\phi_{0}\ll 1$, and take only the leading order terms. This is realistic since it is always expected that the cosmological influence on local systems are minimal. Expression \eqref{erad} can be then simplified, reading
\begin{equation}
\frac{dE}{dt}= \frac{\lambda G^3M^{2}}{2a^{2}(1-e^{2})^{2}c^{5}}\left(f+4M e\cos{\theta}\right)^{2}e^{2}\sin^2{\theta}\,\dot{\theta}^2\,,
\end{equation}
where $f$ is given by
\begin{equation}
f=\frac{5m_{1}m_{2}}{M}- 4M.
\end{equation}
Averaging the energy loss over an orbital period $T$, where
\begin{equation}\label{period}
T=\frac{2\pi a^{3/2}}{\sqrt{GM}}\,,
\end{equation}
we have,
\begin{align}
	\left\langle\frac{dE}{dt}\right\rangle&=\frac{1}{T} \int_{0}^{T}{ \frac{dE}{dt}\,dt}= \frac{1}{T} \int_{0}^{2\pi}{ \frac{dE}{dt} \frac{d\theta}{\dot{\theta}}}
\end{align}
The above integral is directly solved, yielding
\begin{equation}\label{em}
\left\langle\frac{dE}{dt}\right\rangle=\frac{\lambda}{4}\frac{G^4 M^{3}}{a^{5} c^{5}}\,F(m_1,m_2,e),
\end{equation}
with
\begin{align}\label{F}
	F(m_1,m_2,e)\equiv& \ e^{2}(1-e^{2})^{-7/2}\bigg[f^{2} \ +\notag\\[1ex]
	& \ + \ \bigg(\dfrac{f^{2}}{4} +4Mf + 4M^{2}\bigg)e^{2} +2M^{2}e^{4}\bigg].
\end{align}
To finish, we derive how this loss of energy changes the orbital period of the system. From \eqref{period}, one gets that
\begin{align}\label{per}
	\frac{\dot{T}}{T}=& \ \frac{3}{2}\left\langle\frac{\dot{a}}{a}\right\rangle=\frac{3\,a}{Gm_1m_2}\left\langle\frac{dE}{dt}\right\rangle\notag\\[1ex]
	=& \ \frac{3\lambda}{4}\frac{G^3 M^{3}}{m_1m_2\,a^{4} c^{5}}\,F(m_1,m_2,e).
\end{align}
The result has the same proportionality with the constants $G$ and $c$, as in GR, but has a rather more involved dependence on the masses and the eccentricity of the orbit.

Note that equation \eqref{per} must be negative, otherwise it would imply that the masses are moving away from each other. In other words, the system would be increasing their energy by the emission of GW, an unrealistic situation. The function $F$ is positive, as it can be verified by
comparison between the term $f^{2}$ and the part involved by the round brackets multiplying $e^{2}$ (the only part that could be negative),
\begin{align}\label{pos}
	f^{2}+\bigg(\dfrac{f^{2}}{4} +4Mf + 4M^{2}\bigg)=& \ \dfrac{m_1m_2}{4M^{2}}\big[32\,\dfrac{(m_{1}^{2}+m_{2}^{2})}{m_1m_2} \ + \notag\\[1ex]
	& + \ (m_1^{2}+m_2^{2}) + 77m_1m_2\big]>0.\notag 
\end{align}
Since $e^{2}<1$ for elliptical orbits, it follows that $F$ is always positive. Thus to guarantee $\dot{T} < 0$, we must have $\lambda<0$.

We can use the data from the so called Hulse and Taylor pulsar, PSR 1913+16, to constraint $\lambda$ demanding that GSG prediction is in agreement with observations. The data comes from the measurements of time-of-arrivals during $35$ years \cite{Weisberg_2016}. They are collected in Table \ref{tab}. 

\begin{table}[t]
	\caption{Data from PSR 1913+16 \cite{Weisberg_2016} adapted to the present notation. The numbers between round brackets indicate the erros in the last digit.}
	\label{tab}
	\begin{tabular*}{\columnwidth}{@{\extracolsep{\fill}}lr@{}}
		\hline
		Parameter (units) & Value\\
		\hline
		$e$ & $0.6171340(4)$\\
		$m_{1}$ (solar masses) & $1.438(1)$ \\
		$m_{2}$ (solar masses) & $1.390(1)$ \\
		$T$ (days) & $0.32997448918(3)$\\
		$\dot{T}$  & $-2.398(4)\times 10^{-12}$\\
		\hline
	\end{tabular*}
\end{table}

First, we rewrite equation \eqref{per} in a more appropriated form to use the numerical values. Using \eqref{period}, one has
\begin{equation}
\dot{T}=\dfrac{3\lambda\pi}{2}\dfrac{G^{5/3}}{c^{5}}\dfrac{M^{5/3}}{m_1m_2}\,F(m_1,m_2,e)\left(\dfrac{T}{2\pi}\right)^{-5/3}.
\end{equation}
Substituting the numerical values, with the appropriated propagation of errors,  we estimate  $\lambda= -1.111 \pm 0.003$ and $\sigma=3.444\pm0.002$ (through condition \eqref{c1}), constraining the parameters entering in the GW energy momentum tensor.

It is worth to note that the orbital parameters of the binary system are extracted from the timing pulsar observations in a theory-independent way, but the determination of the masses of the pulsar and its companion are model dependent \cite{Damour_2009}. The mass values in Table \ref{tab} are from GR but its usage here is reasonable due to the satisfactory agreement of GSG in the Solar System tests at the post-Newtonian level. However, any modification on these values will lead to a distinct estimation of $\lambda$ but not an invalidation of GSG by pulsars data.


\section{Concluding remarks}
We have presented a discussion on gravitational waves (GW) in the context of the geometric scalar gravity (GSG), a class of theories describing the effects of gravity as a consequence of a modification of spacetime metric in terms of a single scalar field. GSG overcomes the serious drawbacks present in all previous attempts to formulate a scalar theory of gravity. Its fundamental idea rests on the proposal that the geometrical structure of the spacetime is described by a disformal transformation of a conformal flat metric. The model analyzed here has already showed several advances within the scalar gravity program, featuring a good representation of the gravitational phenomena both in the solar system domains as well as in cosmology.

Initially it was shown the procedure used to construct the weak field limit in GSG considering an expansion of the scalar field over a background cosmological solution. Within this approximation scheme the scalar dynamical equations assumes oscillatory solutions that represent GW in the spacetime structure propagating with light velocity, which is in agreement with recent data from GW and gamma-ray burst detections from the merge of a binary neutron star system.

An important distinction appears in the polarization states of the waves, which is characterized by the presence of a longitudinal mode in GSG. Within the $E(2)$-classification of gravitational theories, GSG is of the type $\emph{II}_{6}$, since $\Psi_{2}\neq0$. This is the most general class, where the detection of all the other five polarization modes depend on the observer. The detection of extra polarization states (or the absence of them) shall be a decisive test to alternative theories of gravity \cite{Capozziello2019}. Model-independent tools that allow to see how polarization modes affect the response function in GW detectors has been recently developed and must be also applied in GSG \cite{Chatziioannou2012}. This procedure should be considered in the future.

It was also discussed how to define an energy-momentum tensor for the linear GW, following a field theoretical point of view. An ambiguity emerges since GSG fundamental equation includes a non trivial interaction between matter/energy and the scalar field, leading to non unique expression for the approximated gravitational energy-momentum tensor. This freedom is encoded in the constant parameter $\lambda$, which has directly influence in the energy-loss rate when emitting gravitational waves. Consequently, GSG prediction for the orbital variation of a binary system can be used to constraint the theory's parameter with observational data coming from PSR 1913+16. This numerical computation was performed using GR mass values as a first estimation since GSG is in agreement with classical tests and should not present strong deviations on these values. It is then expected that, after analyzing the so called post-Keplerian parametrization of the theory to extract the mass values of a binary system according to GSG,\footnote{A phenomenological parametrization for binary pulsars introduced by T. Damour \cite{Damour:1991rd}, where the Keplerian and post-Keplerian parameters can be read off.} the theory can be more properly constrained. This task will be addressed in a future work.



\begin{acknowledgements}
	I wish to thank J.C. Fabris and T.R.P. Caram\^es for dedicated reviews of this work. This research is supported by FAPES and CAPES through the PROFIX program.
\end{acknowledgements}

\appendix

\section{The near zone scalar field}\label{apt}
The linearized dynamical equation of GSG is a traditional wave equation, which posses some properties depending whether $R$ (the point where the field is being calculated) is larger or smaller than the typical wavelength $\lambda$ of the solution \cite{PoissonWill}. In the \emph{wave zone}, where $R\gg \lambda$, the difference between the retarded time $t_R$ and $t$ is large, so the time derivative of the field is comparable to the spatial derivative. This is the region where the radiation effects are influent in determining the metric. On the other hand, in the region where $R\ll \lambda$, called \emph{near zone}, the difference between the $t_{R}$ and $t$ are small and the time derivatives becomes irrelevant in front of the spatial derivatives. 

The near zone region is covered by the post-Newtonian approximation of the gravitational field, expanding it in orders of $v/c$, where $v$ is the typical velocities of the source's components, and considering also slow motion. This is the approximation required for the scalar field when integrating the wave equation. Once the scalar field aways appears multiplied by $T^{\mu\nu}$ in the integrand, we only need to know its leading order, i.e. its Newtonian approximation. Thus, equation \eqref{gensol} reduces to
\begin{equation}\label{sol1}
\Phi_N(t,\vec{x})=-\,\frac{\kappa}{4\pi} \frac{\alpha_{0}^{3/2}}{(3-\alpha_{0})}\,\int\dfrac{T^{00}}{|\vec{x}- \vec{x}'|}\,d^{3}x',
\end{equation}
where only the zeroth-order terms is considered in the above integral.

From \eqref {hs} and \eqref{pertm}, and using a multipole expansion, one can easily sees that the metric assumes the form 
\begin{equation}
q_{00}= 1- \frac{2GM}{c^{2}R} + {\cal O}\left(\frac{v^{4}}{c^{4}}\right)\,,
\end{equation}
where 
\begin{equation}
M=\frac{1}{c^{2}}\int T^{00}d^{3}x
\end{equation}
and $G$ attends for the Newton's gravitational constant as measured today and it is a redefinition of the theory's coupling constant,
\begin{equation}\label{gtilda}
G= -\frac{\kappa \,c^{4}\,\alpha_{0}^{5/2}}{2\pi(3-\alpha_{0})^{2}}\,.
\end{equation}
Note that the dependence of $G$ with the cosmological background field implies its change as a result of the evolution of the universe. This effect has not been evident in the previous analysis of GSG's newtonian limit since the cosmological influence was neglect in those works \cite{Bittencourt:2016smd,Novello:2012wr}. We will not discuss its implications in the present work, but it certainly shows the importance of take into account the cosmological background when analyzing the Newtonian and post-Newtonian limits of GSG.

\section{More detailed calculations}\label{b}
In this section we aim to be more clear on the calculation of the integrals of the quantities appearing in expressions \eqref{tl}, \eqref{el} and \eqref{cl}. We start with the linearized conservation law, $\p_\mu T^{\mu\nu}=0$\,, from where is possible to derive the following expressions,
\begin{align}
	\dfrac{d}{dt}\int T^{00}\,d^3x&=0\,,\\[1ex]
	\int T^{ij}\,d^3x&=\frac{1}{2c^2}\,\p_t^2\int T^{00}\,x^i x^j\,d^3x\equiv \dfrac{\ddot{I}^{ij}}{2c^{2}}\,,
\end{align}
where $I^{ij}$ represents the second momenta of mass distribution. Only the trace of the quadrupole moment enters in the field equation and it is directly calculated,
\begin{equation}
I=\int \overset{\sc (0)}{T}\,^{00}\,r^2 d^3x=\sum_{n}\,m_nc^2 r_n^2=\frac{c^2r^2}{M}\,m_1 m_2\,.
\end{equation}
The time derivatives can be now easily calculated.

The remaining integrals does contain the newtonian limit of the scalar field explicitly. For the specific case of binary system as a source, the solution \eqref{sol1} becomes
\begin{equation}
\Phi_N(t,\vec{x})=-\frac{G}{c^2}\frac{(\alpha_{0}-3)}{2\alpha_{0}}\,\sum_{p}\,\frac{m_p}{|\vec{x}-\vec{x}_p|}\,.
\end{equation}
When calculating the Newtonian gravitational potential in the position of one of the particles of the system we have to neglect the infinity self potential, thus
\begin{equation}\label{phin}
\Phi_N(t,\vec{x}_n)=-\frac{G}{c^2}\frac{(\alpha_{0}-3)}{2\alpha_{0}}\,\sum_{p\neq n}\,\frac{m_p}{|\vec{x}_n-\vec{x}_p|}\,,
\end{equation}
where the summation above is taken excluding terms when $p=n$. This can be interpreted as a mass renormalization \cite{straumann2013general}. Using this we can integrate expression \eqref{tl} to give
\begin{equation}
\int \Phi_N T^{00}\,d^3x\a -\frac{(\alpha_{0}-3)}{\alpha_{0}} \ \frac{Gm_1m_2}{r}\,.
\end{equation}

For the remaining integrals it is needed the derivatives of the Newtonian scalar field, namely
\begin{align}
	&\p_0 \Phi_N(t,\vec{x}_n)=-\frac{G}{c^3}\frac{(\alpha_{0}-3)}{2\alpha_{0}}\,\sum_{p\neq n}\,m_p\,\frac{(\vec{r}_{np}\cdot \vec{v}_p)}{r_{np}^3}\,,\\[2ex]
	& \p_i \Phi_N(t,\vec{x}_n)=\frac{G}{c^2}\frac{(\alpha_{0}-3)}{2\alpha_{0}}\,\sum_{p\neq n}\,m_p\,\frac{(x^j_n - x_p^j)\delta_{ji}}{r_{np}^3}\,,
\end{align}
and the kinect term,
\begin{equation}
w_N(t,\vec{x}_n)= -\frac{G^{2}}{c^4}\frac{(\alpha_{0}-3)^{2}}{4\alpha_{0}^{2}}\,\sum_{p,q\neq n}\, m_p m_q \dfrac{(\vec{r}_{np}\cdot \vec{r}_{nq})}{r_{np}^3\,r_{nq}^3}\,.\label{w}
\end{equation}
In the above expressions $\vec{v}_{p}=\dot{\vec{r}}_{p}$, \ $\vec{r}_{np}=\vec{x}_n - \vec{x}_p$, \ and \ $r_{np}=|\vec{r}_{np}|$. The sub-indexes $(p,q,n)$ are summation indices assuming the values $1$ or $2$. The upper-indexes $(i,j,k)$ refer to the usual components of a three-vector and they run from $1$ to $3$.


Let us calculate one of the integrals explicitly,
\begin{multline}
	\int\frac{T^{00}(\p_0 \Phi_N)^2d^{3}x}{w_N} =  \frac{G^{2}}{c^4}\frac{(\alpha_{0}-3)^{2}}{4\alpha_0^{2}}\times \\[1ex]
	\ \times \ \sum_{n}\sum_{p,q\neq n}\,\frac{m_n m_p m_q}{w_N(t,\vec{x}_n)}\,\frac{(\vec{r}_{np}\cdot \vec{v}_p)}{r_{np}^3}\frac{(\vec{r}_{nq}\cdot \vec{v}_q)}{r_{nq}^3},
\end{multline}
where the symbol $\Sigma_{p,q\neq n}$ means the product of two summations, one in $p$ and other in $q$, with both never assuming the value of $n$. Using that
\begin{equation}
w_N(t,\vec{x}_{1(2)})=- \frac{(\alpha_{0}-3)^{2}G^{2}}{4\alpha_0^{2}c^{4}}\frac{m_{2(1)}^2}{r^4},
\end{equation}
and
\begin{equation}
\vec{r}\cdot \vec{v}_n=r\,\dot{r}_n,
\end{equation}
we have
\begin{align}
	\int\frac{T^{00}(\p_0 \Phi_N)^2}{w_N}\,d^{3}x =&-m_1\,\dot{r}_2^2-m_2\,\dot{r}_1^2 \notag\\[1ex]
	& \ - \ \frac{\dot{r}^2}{M^2}\,\left(m_1^3+m_2^3\right)\,.
\end{align}
In the last equality we used the relations \eqref{rs}. The procedure is similar for the other integrals and, paying attention that $\vec{r}_{21}=- \ \vec{r}$, it follows
\begin{align}
	&\int \frac{T^{0i}\p_0 \Phi_N\,\p_i \Phi_N}{w_N}\,d^{3}x=- \ \frac{\dot{r}^2}{M}\,m_1m_2\,,\\[1ex]
	&\int \frac{T^{ij}\p_i \Phi_N \p_j \Phi_N}{w_N}\,d^{3}x=- \ \frac{\dot{r}^2}{M}\,m_1m_2\,.
\end{align}
Putting all these terms together, following \eqref{el}, we obtain the relation,
\begin{equation}\label{int.el}
\int E\,d^{3}x\a - M\dot{r}^{2}
\end{equation}

In the integral of \eqref{cl} the following terms will appear,
\begin{align}
	&\int\dfrac{\Phi_N}{w_N}\left(T^{00}\p_0\Phi_N+T^{0i}\p_i\Phi_N \right)d^{3}x\a cMr\dot{r},\\[2ex]
	&\int\dfrac{T^{00}\p_0\Phi_N+T^{0i}\p_i\Phi_N}{w_N}\,d^{3}x\a - \dfrac{\alpha_0}{\alpha_0-3} \dfrac{cm_1m_2}{GM^{2}}\,\Big(r^{2}\dot{r}^{3} \ + \notag\\[1ex]
	& \hspace{3cm} + \ r^{4}\dot{r}\dot{\theta}^{2}\Big) - \ c\dfrac{(m_1^{2}+m_2^{2})}{M}\,r\dot{r},\\[2ex]
	&\int \dfrac{E\p_0\Phi_N}{w_n^{2}}\,d^{3}x\a \dfrac{8\alpha_0^{2}}{\alpha_0-3}\,\dfrac{c}{G}\,\dfrac{(m_1^{5}+m_2^{5})}{m_1m_2M^{3}}\,r^{2}\dot{r}^{3},
\end{align}
and, with a time derivative, we obtain the last terms remaining to get the expression \eqref{phi}.

\end{sloppypar}

\end{document}